\documentstyle[12pt]{article}

\catcode`\@=11
\long\def\@makefntext#1{
\protect\noindent \hbox to 3.2pt {\hskip-.9pt

$^{{\ninerm\@thefnmark}}$\hfil}#1\hfill}

 \def\@makefnmark{\hbox to 0pt{$^{\@thefnmark}$\hss}}

\def\ps@myheadings{\let\@mkboth\@gobbletwo
\def\@oddhead{\hbox{}
\rightmark\hfil\ninerm\thepage}

\def\@oddfoot{}\def\@evenhead{\ninerm\thepage\hfil
\leftmark\hbox{}}\def\@evenfoot{}
\def\sectionmark##1{}\def\subsectionmark##1{}}

\newcounter{sectionc}\newcounter{subsectionc}
\newcounter{subsubsectionc}
\renewcommand{\section}[1]{\vspace{0.6cm}
\addtocounter{sectionc}{1}

\setcounter{subsectionc}{0}
\setcounter{subsubsectionc}{0}\noindent

         {\bf\thesectionc. #1}\par\vspace{0.4cm}}
\renewcommand{\subsection}[1]
{\vspace{0.6cm}\addtocounter{subsectionc}{1}

         \setcounter{subsubsectionc}{0}\noindent

         {\it\thesectionc.\thesubsectionc.#1}\par
\vspace{0.4cm}} \renewcommand{\subsubsection}[1]
{\vspace{0.6cm}\addtocounter{subsubsectionc}{1}
         \noindent
{\rm\thesectionc.\thesubsectionc.\thesubsubsectionc.#1}
\par\vspace{0.4cm}}

\newcounter{appendixc}
\newcounter{subappendixc}[appendixc]
\newcounter{subsubappendixc}[subappendixc]

\renewcommand{\appendix}[1] {\vspace{0.6cm}
        \refstepcounter{appendixc}
        \setcounter{figure}{0}
        \setcounter{table}{0}
        \setcounter{equation}{0}

\renewcommand{\thefigure}{\Alph{appendixc}.\arabic{figure}}
  \renewcommand{\thetable}{\Alph{appendixc}.\arabic{table}}
  \renewcommand{\theappendixc}{\Alph{appendixc}}

\renewcommand{\theequation}{\Alph{appendixc}.\arabic{equation}}
\noindent{\bf Appendix \theappendixc #1}\par\vspace{0.4cm}}

\renewenvironment{thebibliography}[1]
         {\begin{list}{\arabic{enumi}.}
         {\usecounter{enumi}\setlength{\parsep}{0pt}
\setlength{\leftmargin 1.25cm}{\rightmargin 0pt}
          \setlength{\itemsep}{0pt} \settowidth
         {\labelwidth}{#1.}\sloppy}}{\end{list}}

\newcounter{itemlistc}
\newcounter{romanlistc}
\newcounter{alphlistc}
\newcounter{arabiclistc}

\newcommand{\fcaption}[1]{
        \refstepcounter{figure}
        \setbox\@tempboxa = \hbox{\tenrm Fig.~\thefigure. #1}
    \ifdim \wd\@tempboxa > 6in
           {\begin{center}
   \parbox{6in}{\tenrm\baselineskip=12pt Fig.~\thefigure. #1}
         \end{center}}
        \else
             {\begin{center}
             {\tenrm Fig.~\thefigure. #1}
              \end{center}}
        \fi}

\newcommand{\tcaption}[1]{
        \refstepcounter{table}
        \setbox\@tempboxa = \hbox{\tenrm Table~\thetable. #1}
    \ifdim \wd\@tempboxa > 6in
           {\begin{center}
      \parbox{6in}{\tenrm\baselineskip=12pt Table~\thetable. #1}
            \end{center}}
        \else
             {\begin{center}
             {\tenrm Table~\thetable. #1}
              \end{center}}
        \fi}

\def\@citex[#1]#2{\if@filesw\immediate\write\@auxout
         {\string\citation{#2}}\fi
\def\@citea{}\@cite{\@for\@citeb:=#2\do
         {\@citea\def\@citea{,}\@ifundefined
         {b@\@citeb}{{\bf ?}\@warning
         {Citation `\@citeb' on page \thepage \space undefined}}
        {\csname b@\@citeb\endcsname}}}{#1}}

\newif\if@cghi
\def\cite{\@cghitrue\@ifnextchar [{\@tempswatrue
         \@citex}{\@tempswafalse\@citex[]}}
\def\citelow{\@cghifalse\@ifnextchar [{\@tempswatrue
         \@citex}{\@tempswafalse\@citex[]}}
\def\@cite#1#2{{$\null^{#1}$\if@tempswa\typeout
     {IJCGA warning: optional citation argument ignored: `#2'}
\fi}}

\def\fnt#1#2{\footnotetext{\kern-.3em
         {$^{\mbox{\sevenrm #1}}$}{#2}}}

 1
\font\twelverm=cmr10  scaled\magstep 1
\font\twelveit=cmti10 scaled\magstep 1

\font\elevenrm=cmr10      scaled\magstephalf

\font\tenrm=cmr10

\font\ninerm=cmr9

\font\verybig=cmbx10 scaled\magstep 2
\begin{document}

\centerline{\verybig THE MANY-ANYON PROBLEM}
\vspace{1.0cm}
\centerline{\twelverm CLAUS MONTONEN}
\vspace{0.6cm}
\centerline{\twelveit Department of Physics, Theory Division}
\baselineskip=15pt
\centerline{\twelveit FIN-00014 University of Helsinki, Finland}
\vspace{0.8cm}
\centerline{\elevenrm Lectures presented at the VI Mexican School
of Particles and Fields,}
\baselineskip=14pt
\centerline{\elevenrm Villahermosa, Tabasco, 3--7 October 1994.}
\centerline{\elevenrm HU-TFT-95-12}
\vfil
\rm\baselineskip=14pt
\section{Introduction}
These lectures deal with a particular problem in the theory of
anyons---particles obeying statistics that is neither Bose--
Einstein or Fermi--Dirac, but something in between. Like so many
other developments in theoretical physics nowadays, the concept
of anyons has so far had disappointingly few applications to
observed phenomena---the fractional quantum Hall effect being
the only candidate at the moment. I sincerely hope that this
typical trait of postmodern theory will not apply to anyons in
all future. Anyway, the subject possesses considerable elegance
and is intellectually rewarding, and thus perhaps worth spending
some time upon.

I will start by explaining how the possibility for exotic
statistics arises, and then concentrate on the problem of how to
describe a many--anyon system. By no means will the treatment
cover all the attempts in this direction, nor will the list of
references exhaust the vast literature on the subject. I have
only quoted works I have directly consulted when preparing the
lectures and these notes. I offer my apologies to those whose
work is not discussed or cited.

My involvment in the problem arose from an attempt to describe
comprehensively the statistical mechanics of anyons---an effort
that eventually led to other things. I wish to thank my
collaborators Masud Chaichian and Ricardo Felipe Gonzalez for
many enlightening discussion. An excellent set of lectures given
at the University of Helsinki in 1993 by Finn Ravndal gave
additional inspiration.

\section{The Symmetry Group Approach to the Quantum Mechanics of
Identical Particles}
The average student of quantum mechanics, when first faced with
a treatment of identical particles, would have encountered an
argument that runs somewhat like this:

Consider a system of $N$ particles described by a Hamiltonian
$$H=H(1,2,...,N),\eqno(2.1)$$
where the label $i$ denotes operators (coordinates, momenta,
spins,...) relating to the $i$th particle. The statement that
the particles are identical is taken to mean that any
conceivable Hamiltonian (2.1) describing a system of such
particles is invariant under a permutation of the operators
relating to different particles: Denote by $\pi$ the permutation
$$\pi = \left(\matrix{1&2&\ldots&N\cr \pi (1)&\pi (2)&\ldots&\pi
(N)\cr}\right)$$
belonging to the group $S_N$ of permutations of $N$ objects,
then the particles are identical if
$$U(\pi )H(1,2,...,N)U^{-1}(\pi )\equiv H(\pi (1),\pi
(2),...,\pi (N))=H(1,2,...,N).\eqno(2.2)$$

If this is the case, then according to the general principles of
quantum mechanics, the eigenstates of $H$ should transform
according to some representation of $S_N$:
$$U(\pi )\vert {\psi}_j \rangle = \sum_{k} \vert {\psi}_k
\rangle D_{kj} (\pi ).\eqno(2.3)$$
Denoting by $\vert 1,2,...,N \rangle $ the eigenstates of a
complete set of commuting one-particle operators, on which the
permutation operators act as follows:
$$U(\pi )\vert 1,2,...,N\rangle = \vert \pi (1),\pi (2),...,\pi
(N)\rangle ,\eqno(2.4)$$
the wavefunctions
$$\psi (1,2,...,N) = \langle 1,2,...,N\vert \psi \rangle
\eqno(2.5)$$
transform in the following way:
$$\langle 1,2,...,N \vert U(\pi ) \vert \psi_j \rangle = \psi_j
(\pi^{-1}(1),\pi^{-1}(2),...,\pi^{-1}(N)) = \sum_k \psi_k
(1,2,...,N) D_{kj} (\pi ),$$
or, by a relabelling of the arguments of the wave functions:
$$\psi_j (1,2,...,N) = \sum_k \psi_k (\pi (1),\pi (2),...,\pi
(N)) D_{kj}(\pi ).\eqno(2.6)$$

A textbook in group theory would tell us that there are exactly
two one-dimensional representations of any $S_N$, $N\geq 2$:

- The trivial representation $D(\pi )=1$. Particles transforming
according to this representation are called bosons and their
wavefunctions are completely symmetric.

- The alternating representation $D(\pi )=(-1)^{\vert \pi \vert
}$, where $\vert \pi \vert$ denotes the number of exchanges
needed to build the permutation (although not unique, this
number is always either even or odd for a given $\pi$).
Particles transforming in this way are called fermions, and
their wavefunctions are completely antisymmetric.

Irreducible representations of higher dimensions do occur for
$N\geq 3$, and the term parastatistics has been introduced to
describe this situation. Parastatistics will not be treated
further in this course, be it either because it does not seem to
occur in nature or because general theorems {\cite{1}} say that
parastatistics can always be replaced by hidden ("colour")
degrees of freedom.

The alternative between Bose or Fermi statistics, which the
above argument led us to, can be expressed in a simple way,
which, however, is an extremely powerful tool for computations.
I am referring to the formalism of second quantization. Let
$\psi^{\dagger}({\bf x})$ be the operator creating a particle at
${\bf x}$, and $\psi ({\bf x})$ the operator annihilating a
particle at ${\bf x}$ (for simplicity we assume that the
particle number is conserved as e.g. in nonrelativistic many-
body theory), and $\vert 0 \rangle$ the vacuum (no particle)
state. Then the particle statistics is all contained in the
algebraic relations (${\lbrack A,B\rbrack}_\mp \equiv AB \mp BA$
$$\lbrack \psi ({\bf x}),\psi ({\bf x'})\rbrack_{\mp} = \lbrack
\psi^{\dagger}({\bf x}),\psi^{\dagger}({\bf x'})\rbrack_{\mp} =
0,\eqno(2.7)$$
$$\lbrack \psi ({\bf x}),\psi^{\dagger}({\bf x'})\rbrack_{\mp} =
\delta ({\bf x} - {\bf x'}),\eqno(2.8)$$
$$\psi ({\bf x})\vert 0 \rangle = 0,\eqno(2.9)$$
where the - sign applies to bosons, the + sign to fermions. The
state with particles localized at ${\bf x}_1,...,{\bf x}_N$ is
represented by the state vector
$$\vert {\bf x}_1,...,{\bf x}_N\rangle = \psi^{\dagger}({\bf
x}_1)\cdots \psi^{\dagger}({\bf x}_N)\vert 0 \rangle
\eqno(2.10)$$
and the relations (2.7) automatically ensure the correct
symmetry properties.

A nice and important fact is that the Bose--Fermi--alternative
looks the same in any representation: Indeed, in Eqs. (2.4) and
(2.5) we did not specify which representation we were using. In
the language of second quantization, we see this as follows:

Let $\lbrace u_n({\bf x})\rbrace$ be a complete set of
orthonormal functions (eigenfunctions of a one--particle
operator):
$$\int d{\bf x} u_n^{\ast}({\bf x})u_m({\bf x}) = \delta_{nm},$$
$$\sum_n u_n({\bf x})u_m^{\ast}({\bf x'}) = \delta ({\bf x} -
{\bf x'}).$$
Expanding $\psi ({\bf x}),\psi^{\dagger} ({\bf x})$:
$$\psi ({\bf x}) = \sum_n a_nu_n({\bf x})$$
$$\psi^{\dagger} ({\bf x}) = \sum_n a_n^{\dagger}u_n^{\ast}({\bf
x}),$$
one easily derives the algebra of the operators
$a_n,a_n^{\dagger}$, which is formally {\it identical} to
(2.7)--(2.8):
$$\lbrack a_n,a_m \rbrack_{\mp} = \lbrack
a_n^{\dagger},a_m^{\dagger} \rbrack_{\mp} = 0$$
$$\lbrack a_n,a_m^{\dagger} \rbrack_{\mp} =
\delta_{nm}.\eqno(2.11)$$

The fact that the second quantization is equally simple in any
representation is of crucial importance e.g. in the relativistic
case, where (asymptotic) states of sharp momenta make sense, but
states of sharp localization do not.

Can more exotic possibilities for the statistics of identical
particles be envisaged? The answer, gleaned at in partial
results for 1+1--dimensional field theories {\cite{2}}, was
definitively shown to be yes by Leinaas and Myrheim in 1977
{\cite{3}}, provided the dimension of space is 1 or 2.
Subsequently {\cite{4}}, the name {\it anyons} (in Spanish:
qualquierones, according to Eduardo Fradkin) was given to
particles obeying such exotic statistics. It is evident that all
parts of the previous reasoning cannot apply to anyons (since
our argument uniquely led to bosons or fermions), but it is of
interest to investigate how much of it can be saved, not least
because of the computational ease of the second quantized
formalism.

As an example, consider generalizing the relations (2.7) to
$$\psi ({\bf x})\psi ({\bf x'}) = \alpha \psi ({\bf x'})\psi
({\bf x}).\eqno(2.12)$$
Exchanging ${\bf x}$ and ${\bf x'}$ we see that consistency
requires $\alpha = \alpha^{-1}$, i.e. $\alpha = \pm 1$. Thus we
are back at the Bose-Fermi alternative. Hence $\alpha$ cannot be
a constant, rather should we take
$$\psi ({\bf x})\psi ({\bf x'}) = \alpha ({\bf x},{\bf x'}) \psi
({\bf x'})\psi ({\bf x}),\eqno(2.13)$$
with $\alpha ({\bf x},{\bf x'}) = \alpha^{-1}({\bf x'},{\bf
x})$. But the form (2.13) is now peculiar to ${\bf x}$-space; in
${\bf p}$-space, e.g., the relation will look completely
different!

\section{How Come Anyons?}
In order to clearly see what new features are implied in the
reasoning leading to anyon statistics, I will here present a
strict, orthodox party line. Like all party lines, it should be
constantly challenged, and ways to overthrow the orthodoxy
should be sought. In this way new discoveries can be made, and
what survives of the orthodoxy will stand on more secure ground.
So here we go:

By identical particles we shall mean the following: Firstly, the
configuration space for $N$ identical particles in $D$--
dimensional Euclidean space is not $({\bf R}^D)^N$, instead
$({\bf x}_1\ldots {\bf x}_N)$ should be identified with $({\bf
x}_{\pi(1)}\ldots {\bf x}_{\pi(N)})$ for any $\pi \in S_N$. The
space obtained after such an identification, denoted by $({\bf
R}^D)^N{\slash}S_N$, has the awkward property of possessing
potentially singular points. The candidates for such points are
the fixed points of the action of $S_N$ on $({\bf R}^D)^N$, i.e.
the {\it diagonal} $\Delta \equiv \lbrace ({\bf x}_1\ldots {\bf
x}_N) \in ({\bf R}^D)^N\vert {\bf x}_i = {\bf x}_j\: \rm for\:at\:
least\:one\:pair \rbrace$. So, to stay clear of trouble we should
remove the diagonal (we could imagine that there is a hard core
interaction between the particles keeping them apart). The first
statement of our dogma is thus that the configuration space of
$N$ identical particles in $D$-dimensional space is
$$M_N^D = {({\bf R}^D)^N - \Delta\over S_N}.\eqno(3.1)$$

Secondly, we will allow as observables only symmetric operators.
This means e.g. that we are not allowed to consider one--
particle operators such as $H_0(i) = {{\bf p}_i^2\slash 2m}$,
although in the symmetry group way of looking at things a
quantity like $\langle H_0(1)\rangle$ makes sense (it is
perfectly calculable), by symmtery, of course, it equals
$\langle H_0(2)\rangle = \cdots = \langle H_0(N)\rangle$.

How can we then introduce the concept of particle statistics,
since by adopting the above dogma we have banished all talk
about "interchanging particles" and the like? The key point is
that the configuration manifold (3.1) is (for $D\geq2$) not
simply connected: There are closed loops in $M_N^D$ that cannot
be continuously shrunk to a point. A simple example makes this
clear. $M_2^2$ can be constructed as follows: The coordinates
${\bf x}_1,{\bf x}_2$ of $({\bf R}^2)^2$ are replaced by the
center of mass coordinate ${\bf X} = {1\over 2}({\bf x}_1 + {\bf
x}_2)$ and the relative coordinate ${\bf x} = {\bf x}_1 - {\bf
x}_2$. Removing the diagonal ${\bf x}_1 = {\bf x}_2$ means
leaving out the origin of the ${\bf x}$--plane. "Modding" by
$S_2$ means identifying ${\bf x}$ and $-{\bf x}$. This we can do
by restricting us to the left half ${\bf x}$--plane $x^1\geq 0$.
On the $x^2$--axis we still have to modify the points $(0,x^2)$
and $(0,-x^2)$, i.e. glue the negative $x^2$--axis to the
positive $x^2$--axis. The resulting construction is the surface
of a cone with the tip (${\bf x} = 0$) excluded:
$$M_2^2 = {\bf R}^2\times \lbrace \rm cone\:without\:
the\:tip \rbrace .$$
Evidently, any closed loop on the mantle of the cone encircling
the tip cannot be shrunk to a point; thus $M_2^2$ is multiply
connected.

Quantum mechanics on multiply connected spaces shows interesting
new features not present in the familiar case when the
configuration space is topologically trivial, and the
possibility of exotic statistics lies hidden in these new
traits. I will present the argument using Feynman's path
integral formulation of quantum mechanics. If you prefer the
Schr{\"o}dinger wave function formulation, I recommend the book
by Morandi~{\cite{5}}.

In Feynman's formulation, the propagator for a configuration
$a\in M_N^D$ at time $t_a$ to develop into a configuration $b$
at $t_b$ is given by the path integral
$$K(b,t_b;a,t_a) = \int_{q(t_a)=a}^{q(t_b)=b}{\cal
D}qe^{iS}.\eqno(3.2)$$
Here $S$ is the action, and the integral runs over all paths
$q(t)$ in $M_N^D$ connecting $a$ to $b$. As $M_N^D$ is multiply
connected, there are paths which cannot be continuously deformed
into each other. All paths which can be deformed into each other
we group into the same {\it homotopy class}. The set of all
paths from $a$ to $b$ is thus divided into homotopy classes, and
we denote the set of all homotopy classes by $\pi (M_N^D,a,b)$.

As was first pointed out by Schulman {\cite{6}}, in this case we
can a priori weight the contributions from different classes
differently:
$$K = \sum_{a\in\pi}\chi(\alpha)K^{\alpha},\eqno(3.3)$$
where the sum runs over all classes, and $K^{\alpha}$ denotes
the integral over all paths in the class ${\alpha}$.

What consistency conditions do the weights $\chi(\alpha)$ have
to satisfy? The answer to this question was given by Laidlaw and
Morette--De Witt {\cite{7}}. The argument is quite subtle, and I
shall only summarize the results here, referring the reader to
the original paper for details.

Firstly, the weights $\chi(\alpha)$ have to be pure phases:
$\vert \chi (\alpha ) \vert = 1$. The reason for this is
basically that when $t_b-t_a \rightarrow 0$, only one class
contributes, and this term has to reproduce $K$ up to a phase.

Secondly, we make the following observation. Let us choose a
fixed point $q_0 \in M_N^D$, and fixed paths $C_a,C_b$ joining
$a$ and $b$ to $q_0$, respectively. Then in each class $\alpha$
there are representatives consisting of:

{\obeylines - the path $C_a$
- a closed loop in $M_N^D$ starting and ending at $q_0$
- the path $C_b$.}

Now the set of all closed loops at $q_0$ falls into classes for
which we can introduce a "multiplication": The product of two
loops $\gamma ,\gamma'$ is the loop formed by first going around
$\gamma$ and then around $\gamma'$. With this multiplication the
set of classes of closed loops at $q_0$ forms a group (the unit
element being the class to which the constant loop staying at
$q_0$ belongs, and the inverse of the class of the loop $\gamma$
being the class to which traversing $\gamma$ in the opposite
direction belongs), the {\it fundamental} or {\it first homotopy
group} $\pi_1(M_N^D)$ (we can drop the reference to $q_0$, since
all the groups at different $q_0$ are isomorphic).

In this way we establish a one-to-one correspondence between
$\pi (M_N^D,a,b)$ and $\pi_1(M_N^D)$ (which by the way is not
unique, since it depends on the choice of $C_a$ and $C_b$!), and
$\alpha$ can be taken as labelling $\pi_1(M_N^D)$.

If $t_c$ is a time intermediate between $t_a$ and $t_b$, the
propagator has to obey
$$K(b,t_b;a,t_a) = \int_{M_N^D} dc
K(b,t_b;c,t_c)K(c,t_c;a,t_a).\eqno(3.4)$$
For the classes this means
$$K^{\gamma}(b,t_b;a,t_a) = \sum_{\alpha , \beta ; \alpha \cdot
\beta = \gamma}\int dc
K^{\beta}(b,t_b;c,t_c)K^{\alpha}(c,t_c;a,t_a). \eqno(3.5)$$
Together, (3.4) and (3.5) imply that
$$\chi (\alpha )\chi (\beta ) = \chi (\gamma = \alpha \cdot
\beta ),\eqno(3.6)$$
i.e. the weights form a unitary, one-diemnsional representation
of $\pi_1(M_N^D)$ {\cite{8}}.

These groups are known:
$$\pi_1(M_N^2) = B_N,\eqno(3.7)$$
the {\it N-string braid group}, whereas
$$\pi_1(M_N^D) = S_N \eqno(3.8)$$
for $D\geq 3$. Both $B_N$ and $S_N$ are generated by $N-1$
generators $\sigma_1\ldots\sigma_{N-1}$, obeying the constraints
$$\sigma_i \sigma_j = \sigma_j \sigma_i , \vert i-j \vert \geq
2, \eqno(3.9)$$
$$\sigma_i \sigma_{i+1} \sigma_i = \sigma_{i+1} \sigma_i
\sigma_{i+1} .\eqno(3.10)$$
The difference between $B_N$ and $S_N$ arises from the fact that
for $S_N$ we require in addition to (3.9) and (3.10)
$$\sigma_i^2 = e, \eqno(3.11)$$
where $e$ is the unit element.

The connection to particle statistics comes through recognizing
that the class of closed loops $\sigma_i$ corresponds to an
interchange of particles $i$ and $i+1$. In the plane ($D=2$)
this can be done in two homotopically inequivalent ways, which
can be represented by the loop where these two particles move on
the circumference of a circle passing through their original
locations either counterclockwise (corresponding to $\sigma_i$)
or clockwise (corresponding to $\sigma_i^{-1}$) interchanging
their places, whereas all other particles stay put. In three or
more dimensions these loops can be deformed into each other e.g.
by rotating the circle around a diameter, i.e. ${\sigma_i =
\sigma_i^{-1}}$ in accordance with (3.11), but not in two
dimensions.

The elements of the group $B_N$ ($S_N$) are formed by taking all
possible products of all possible powers (positive and negative)
of the generators $\sigma_i$, taking into account the
constraints (3.9), (3.10) ((3.9), (3.10), (3.11)). $B_N$ is a
group of infinitely many elements, but the inclusion of the
powerful constraints (3.11) reduces the number of elements of
$S_N$ to $N!$.

Our general result, Eq. (3.6), instructs us to look for unitary,
one-dimensional representations of $B_N$ or $S_N$. Posing
$$\chi (\sigma_i ) = e^{i\phi_i},$$
we see that Eq. (3.10) requires $\phi_i = \phi_{i+1}$, i.e. all
phases are equal. It is customary to write
$$\chi (\sigma_i ) = e^{-i\nu \pi}$$
$$\chi (\sigma_i^{-1}) = e^{i\nu \pi},\eqno(3.12)$$
where $\nu \in [0,2)$ is the {\it statistical parameter}.

In three or more dimensions, Eq. (3.11) requires
$\chi(\sigma_i)^2 = 1$, i.e. $\nu = 0,1$ are the only
possibilities (we have in fact derived the result on the one-
dimensional representations of $S_N$ mentioned in section 2!).
But for $D=2$ there is no restriction on $\nu$, and anyon
statistics is possible.

This, then, is one version of the accepted orthodoxy on
unorthodox statistics. Parts of it can be challenged. For
instance, one might ask what happens if we do not remove the
diagonal but work with $({\bf R}^D)^N/S_N$ as the configuration
space (which then is no longer a manifold, but rather an
"orbifold"). As is evident from our example $M_2^2$, this will
change the fundamental group of the configuration space, and our
previous argument breaks down. In simple cases, at least, it
seems that Hamiltonians on $M_N^2$ can be extended to self-
adjoint operators on ${\bf R}^{2N}/S_N$ ("colliding anyons") in
many ways {\cite{9}}. What this means is still uncertain.

Although we in the sequel will be exclusively concerned with
exotic statistics in two-dimensional space, let us briefly stop
to consider what happens in one dimension. In this case the
configuration space is simply connected:
$$\pi_1(M_N^1) = 0,$$
and our previous argument seems to imply that no statistics is
possible. In a certain sense this is true: To exchange two
particles on a line, they have to be moved past each other, and
what then happens depends on any contact interaction between the
particles. In other words, there is a possibility of introducing
statistics through the boundary conditions to be imposed at the
edges of $M_N^1$. The following example, taken from Leinaas and
Myrheim {\cite{3}}, illustrates this point:

Take two particles on a line, with coordinates $x_1$ and $x_2$.
$M_2^1$ is e.g. the region to the right of the diagonal $x_1 =
x_2$ of the $(x_1,x_2)$-plane, and we have to decide what
boundary conditions to impose on the diagonal. The normal
derivative of the wave function on the boundary is the partial
derivative with respect to the relative coordinate $x = x_1 -
x_2$, and a natural condition on the wave function would be that
the probability current vanishes at the boundary (no probability
"flows out of" $M_2^1$):
$$j_n \propto i(\psi^* {\partial \psi \over \partial x} - \psi
{\partial \psi^* \over \partial x})\vert_{x=0} = 0.$$
The solution to this condition is that
$${\partial \psi \over \partial x}\vert_{x=0} = \eta \psi
(x=0)$$
for any real $\eta$. Choosing $\eta = 0$ allows $\psi$ to be
extended to an even function of $x$ in the whole plane, i.e.
Bose statistics; if $\eta = \pm \infty, \psi (x=0) = 0$, and
$\psi$ can be extended into an antisymmetric function, i.e. we
get Fermi statistics. For any other value of $\eta$ we get
statistics intermediate between Bose and Fermi; i.e. what
corresponds to anyons in $D=1$.

\section{The Transmutation of Statistics into a Topological
Interaction}
A direct attack on the anyon problem using the boundary
conditions on the wave function of $N$ anyons implied by the
propagator (3.3) with weights (3.12) is easy for $N=2$
{\cite{3}}, but already for $N=3$ it becomes a very difficult
problem, and significant progress in solving the three-anyon
problem was achieved only very recently {\cite{10}}. Another
approach, whereby the exotic statistics is transformed into a
peculiar interaction between ordinary bosons or fermions, has
become more popular. In this section, we shall derive this
statistical interaction, following Wu {\cite{8}}, and in the
next section we shall show that this same statistical interaction
can be generated by introducing a gauge field with a very
special kinetic term, the famous Chern--Simons term.

Define
$$z_{ij} = (x_j^1-x_i^1) + i(x_j^2-x_i^2) = \vert z_{ij} \vert
e^{i\phi_{ij}}.\eqno(4.1)$$
As a loop representing the class $\sigma_i$ we can take the loop
in $M_N^2$ where particles $i$ and $i+1$ exchange their places
by rotating counterclockwise through $\pi$ around the center
point of the line joining their original positions, whereas all
other particles stay where they are. Their relative polar angle
changes by $\pi$:
$$\Delta \phi_{i,i+1} = \int_{t_a}^{t_b} dt {d \over
dt}\phi_{i,i+1} (t) = \pi,\eqno(4.2)$$
i.e. we can write
$$\chi (\sigma_i) = e^{-i\nu \pi} = exp(-i\nu \int_{t_a}^{t_b}
dt {d \over dt} \phi_{i,i+1} ).\eqno(4.3)$$
The changes in all other relative polar angles add up to zero,
because $\Delta \phi_{jk} = 0$, $j,k \not= i,i+1$, and $\Delta
\phi_{ij} + \Delta \phi_{i+1,j} = 0, j \not= i,i+1$. For the
inverse $\sigma_i^{-1}$, take a clockwise rotation $\Delta
\phi_{i,i+1} = -\pi$, and
$$\chi (\sigma_i^{-1}) = e^{+i\nu \pi} = exp(-i\nu
\int_{t_a}^{t_b} dt {d \over dt} \phi_{i,i+1} ) \eqno(4.4)$$
again. Since any loop can be built up out of products of the
generators $\sigma_i$, we see that
$$\chi (\alpha ) = exp(-i\nu \int_{t_a}^{t_b} dt {d \over dt}
\sum_{i < j} \phi_{ij}(t)). \eqno(4.5)$$

Thus, supposing the dynamics of the anyons is described by a
Lagrangian $L$, the propagator can be written
$$K = \sum_{\alpha} \chi (\alpha ) \int_{q(t)\in \alpha} {\cal
D}q e^{i\int_{t_a}^{t_b} dt L} = \sum_{\alpha} \int_{q(t)\in
\alpha} {\cal D}q e^{i\int_{t_a}^{t_b} dt L_{eff}} = \int_{all\:
paths} {\cal D}q e^{i\int_{t_a}^{t_b} dt L_{eff}}, \eqno(4.6)$$
which is the path integral of boson particles with a Lagrangian
$$L_{eff} = L - \nu \sum_{i < j} {d \over dt} \phi_{ij} (t).
\eqno(4.7)$$

The last term, the {\it statistical interaction}, is a total
derivative (and thus a "topological term") and will not affect
the equations of motion of the system. Its only role is to
provide the correct statistical weight factors in the
propagator.

Let us look more closely at the case of free (nonrelativistic)
anyons. Then
$$L_{eff} = \sum_{i=1}^N {m \over 2}{\dot {\bf x}}_i^2 - \nu
\sum_{i < j} {\dot \phi}_{ij}. \eqno(4.8)$$
The time--derivative of the polar angle can be written
$${\dot \phi}_{ij} = ({\dot {\bf x}}_i \cdot \nabla_i + {\dot
{\bf x}}_j \cdot \nabla_j )\phi_{ij} = ({\dot {\bf x}}_i - {\dot
{\bf x}}_j)\cdot \nabla_i \phi_{ij} = ({\dot {\bf x}}_j - {\dot
{\bf x}}_i)\cdot \nabla_j \phi_{ij}.$$
Thus the canonical momentum corresponding to ${\bf x}_i$ is
$${\bf p}_i = {\partial L_{eff} \over \partial {\dot {\bf x}}_i}
= m{\dot {\bf x}}_i - \nu \nabla_i \sum_{j\not= i} \phi_{ij}
\equiv m{\dot {\bf x}}_i + {\bf a}_i, \eqno(4.9)$$
where, by abusing three--dimensional notation,
$${\bf a}_i = \nu \sum_{j\not= i} {{\bf e}_3 \times ({\bf x}_i -
{\bf x}_j) \over \vert {\bf x}_i - {\bf x}_j \vert^2}.
\eqno(4.10)$$
(${\bf e}_3$ is a unit vector perpendicular to the plane where
the particles move.)

The Hamiltonian of the system is
$$H = \sum_{i=1}^{N} {\dot {\bf x}}_i \cdot {\bf p}_i - L_{eff}
= {1 \over 2m} \sum_{i=1}^N ({\bf p}_i - {\bf a}_i)^2,
\eqno(4.11)$$
which is of the same form as the minimally coupled Hamiltonian
for $N$ particles moving in an abelian gauge field described by
the vector potential ${\bf A}({\bf x}_i) = {\bf a}_i$. Indeed,
${\bf a}_i$ has been given the name the {\it statistical gauge
field}.

{}From Eq. (4.9) we see that in gauge theory language the
statistical gauge field is a pure gauge, and can thus be removed
by a gauge transformation:
$$\tilde \psi = e^{i\nu \sum_{i< j} \phi_{ij}} \psi_{\rm
Bose}(z_1,\ldots,z_N,z_1^*,\ldots,z_N^*) = \prod_{i< j}
(z_{ij})^{\nu} \Phi_{\rm
Bose}(z_1,\ldots,z_N,z_1^*,\ldots,z_N^*);\eqno (4.12)$$
$(z_k = x_k^1 + ix_k^2, z_{ij} = z_i - z_j)$
$$\tilde H = e^{i\nu \sum_{i< j} \phi_{ij}} H e^{-i\nu
\sum_{i< j} \phi_{ij}} = \sum_{i=1}^N {{\bf p}_i^2 \over 2m}.
\eqno(4.13)$$
The transformed wave functions are not single--valued as
functions on $M_N^2$:
$$\tilde \psi ({\bf x}_1,\ldots,{\bf x}_k,\ldots,{\bf
x}_l,\ldots ,{\bf x}_N) = e^{\pm i \nu \pi} \tilde \psi ({\bf
x}_1,\ldots ,{\bf x}_l,\ldots ,{\bf x}_k,\ldots ,{\bf x}_N),$$
since $\phi_{kl} = \phi_{lk} \pm \pi$ (depending on which way we
braid, i.e. interchange ${\bf x}_k$ and ${\bf x}_l$). Rather,
$\tilde \psi$ is a proper function on the universal covering
space of $M_N^2$, and interchanging particles takes us from one
fundamental domain to another {\cite {5}}. Of course, the gauge
transformation (4.12), (4.13) has taken us back to our starting
point: Eq. (4.12) is the form of the wave function implied by
the propagator (3.3).

\section{The Chern--Simons Action and Anyon Statistics {\cite
{11}}}
The (abelian) Chern--Simons (CS) action is
$$S_{CS} = \int d^3x {\cal L}_{CS} = {\kappa \over 2} \int d^3x
{\epsilon}^{\alpha \beta \gamma} A_{\alpha} {\partial}_{\beta}
A_{\gamma} \eqno(5.1)$$
(${\epsilon}_{012} = {\epsilon}^{012} = +1$). The action (5.1)
is invariant under the $U(1)$ gauge transformation $A_{\mu}
\rightarrow A_{\mu} + {\partial}_{\mu} \Lambda$, since ${\cal
L}_{CS}$ changes only by a total derivative. Let us couple the
CS vector potential to a current $j^{\mu}$ describing $N$ point
particles:
$$j^{\mu}(x) = g \sum_{n=1}^N v_n^{\mu} {\delta}_2({\bf x} -
{\bf x}_n(t)),\eqno(5.2)$$
where the 3-velocity $v^{\mu} = (1,{\bf v})$, and g is the "CS--
charge". The coupling is
$$S_{int} = - \int d^3x j^{\mu}(x) A_{\mu}(x) \eqno(5.3)$$
($g_{\mu \nu} = {\rm diag}(1,-1,-1)$). Let the particles move
nonrelativistically:
$$S_{matter} = \int dt \sum_{n=1}^N {m \over 2} {\bf v}_n^2.
\eqno(5.4)$$
The total action of the model we shall study in this section is
then
$$S = S_{matter} + S_{CS} + S_{int}.\eqno(5.5)$$

The equations of motion can be straightforwardly derived.
Introducing the CS-field strength tensor
$$F_{\mu \nu} = {\partial}_{\mu}A_{\nu} -
{\partial}_{\nu}A_{\mu} $$
with components $E^i \equiv F_{0i}$ (CS--electric field) and $B
\equiv F_{21}$ (CS-magnetic field; in 2+1 dimensions the
magnetic field is a pseudoscalar), we can write the Lorentz
force equations for the particles:
$$m{\dot v}_n^i = g(E^i(t,{\bf x}_n) + {\epsilon}^{ij} v_n^j
B(t,{\bf x}_n)) \eqno(5.6)$$
(${\epsilon}^{12} = {\epsilon}_{12} = +1$). Varying the action
with respect to $A_{\mu}$ we get the field equations
$$j^{\mu} = {\kappa \over 2} {\epsilon}^{\mu \nu \rho} F_{\nu
\rho}, \eqno(5.7)$$
i.e.
$$E^i = {1 \over \kappa} {\epsilon}^{ij} j_j, \eqno(5.8)$$
$$B = -{1 \over \kappa} j^0 \equiv -{1 \over \kappa} \rho.
\eqno(5.9)$$

Since the CS action is of first order in the derivatives of the
fields, the field equations simply express the fields as
functions of the sources and allow the complete elimination of
the fields from the equations of motion. Inserting Eqs (5.8) and
(5.9) in (5.6) we get
$$m{\dot v}_n^i = {g^2 \over \kappa} \sum_{m=1}^N
{\epsilon}^{ij} (v_m^j(t) -v_n^j(t)) {\delta}_2({\bf x}_n(t) -
{\bf x}_m(t)). \eqno(5.10)$$
We see that the Lorentz force vanishes almost everywhere, so
that our system describes free particles. The Chern--Simons term
is not without consequences, however. This is most clearly seen
in the Hamiltonian picture. By following the standard canonical
procedure we derive the Hamiltonian
$$H = \sum_{n=1}^N {1 \over 2m} ({\bf p}_n - g{\bf A}({\bf
x}_n,t))^2 + \int d^2x A_0({\bf x})(\kappa B({\bf x}) + \rho
({\bf x})). \eqno(5.11)$$
We still have the freedom to choose a suitable gauge. The clever
choice is:
$$A_0 = 0, {\partial}_i A^i = 0. \eqno(5.12)$$
The latter condition allows us to solve Eq. (5.9) (which in the
Hamiltonian formulation is to be imposed as a constraint, like
Gauss' law in the Hamiltonian formulation of Maxwell
electrodynamics) uniquely for the CS vector potential:
$$A^i({\bf x}) = {1 \over 2\pi \kappa} \int d^2x'
{\epsilon}^{ij} {x^j - x'^j \over \vert {\bf x} - {\bf x'}
\vert^2} \rho ({\bf x'}) = {g \over 2 \pi \kappa} \sum_{m=1}^N
{\epsilon}^{ij} {x^j - x_m^j \over \vert {\bf x} - {\bf x'}
\vert^2}. \eqno(5.13)$$

Substituting Eq. (5.13) into Eq. (5.11), where the last term is
zero in the gauge (5.12), and dropping the troublesome divergent
$n=m$ terms (it has been argued that this can be justified
through suitable regularization), we arrive at exactly the
statistical interaction (4.11), if we identify the statistical
parameter as
$$\nu = {g^2 \over 2 \pi \kappa}. \eqno(5.14)$$

Thus, the Chern--Simons field minimally coupled to matter
particles generates anyon statistics! This important observation
points to a way of constructing a full-fledged field theory of
anyons, a topic to which we shall now turn.

\section{Nonrelativistic Chern--Simons (--Maxwell) Field Theory}
Quantum field theory remains the preferred vehicle for many-body
quantum theory. In the relativistic case, when particle numbers
are not conserved, it is practically the only available
alternative, but even in the nonrelativistic case it is
computationally superior to its rivals. Thus it is natural to
attempt to base a many-anyon theory on a quantum field theory of
anyons.

However, we do not yet know what such a field theory should look
like. In view of the results of the preceding section, according
to which the minimal interaction of particles with a Chern--
Simons field induces anyon statistics for the particles, it is
tempting to start from a theory of a matter field in interaction
with a Chern--Simons field.

Sticking to the nonrelativistic case, we might try
$${\cal L}_0 = i{\psi}^{\dagger}D_0 \psi + {1 \over
2m}{\psi}^{\dagger}{\bf D}^2 \psi + {\kappa \over 2}
{\epsilon}^{\alpha \beta \gamma} A_{\alpha} {\partial}_{\beta}
A_{\gamma}, \eqno(6.1)$$
where $\psi$ is a boson field, and $D_{\mu} = {\partial}_{\mu} +
igA_{\mu}$. To the Lagrangian (6.1) we might optionally add
further terms, like a Maxwell term
$${\cal L}_M = -{1 \over 4e^2} F_{\mu \nu} F^{\mu \nu},
\eqno(6.2)$$
or a contact interaction between the $\psi$-particles:
$${\cal L}_1 = {\lambda \over 2} :(\psi^{\dagger} \psi )^2:.
\eqno(6.3)$$
The advantage in adding a Maxwell term lies in making the theory
more regular in that $A_{\mu}$ then becomes a physical degree of
freedom (the transverse part of $A_{\mu}$ describes in that case
a massive "photon" of mass $m_{\gamma} = e^2\kappa$). The
contact term (6.3) again brings in new interesting features. The
classical theory based on ${\cal L}_0 + {\cal L}_1$ has soliton
solutions {\cite{13,14}}. L{\"u}scher has studied the theory
${\cal L}_0 + {\cal L}_M$ {\cite{15}}. We shall here see what
follows from ${\cal L}_0$ alone, mainly following
{\cite{12,13}}. In order not to get bogged down in lots of
detail, we will proceed briskly ahead with a heuristic account
of the main line of argument and refer to the original papers
for a more careful treatment (see also {\cite{16,17}}).

We are interested in a second quantized theory of anyons. Under
the assumption that Eq. (6.1) does describe anyons, our strategy
will be to eliminate the Chern--Simons field through an
appropriate gauge transformation and to investigate the
properties of the transformed field operators.

In the pure Chern--Simons case, the equations of motion again
connect the field strength operators to the particle field
operators:
$${\epsilon}^{\alpha \beta \gamma}F_{\beta \gamma} = {2g \over
\kappa} j^{\alpha}, \eqno(6.4)$$
where now
$$j^0 \equiv \rho = {\psi}^{\dagger} \psi \eqno(6.5)$$
$$j^k = {i \over 2m}({\psi}^{\dagger} D^k \psi -
(D^k\psi)^{\dagger} \psi). \eqno(6.6)$$
In particular, Eq. (6.4) says for $\alpha = 0$:
$$B = F_{21} = -{g \over \kappa} \rho. \eqno(6.7)$$
In a transverse gauge, $\nabla \cdot {\bf A} = 0$, this equation
can be solved for the vector potential ${\bf A}$:
$$A^i({\bf x},t) = {\epsilon}^{ij} {\partial \over \partial x^j}
({g \over \kappa} \int d^2y G({\bf x} - {\bf y}) \rho ({\bf
y})). \eqno(6.8)$$
Here $G({\bf x})$ is the two-dimensional Green function
$${\nabla}^2 G({\bf x}) = {\delta}_2 ({\bf x}),$$
i.e.
$$G({\bf x}) = {1 \over 2 \pi} \log(\mu \vert {\bf x} \vert),$$
where $\mu$ is an arbitrary scale. Introducing again the polar
angle $\phi({\bf x})$ through
$$z({\bf x}) \equiv x^1 + ix^2 = \vert z \vert e^{i\phi ({\bf
x})}, \eqno(6.9)$$
the Cauchy--Riemann equations for $f(z) = \log z$ read
$${\epsilon}^{ij} {\partial \over \partial x^j} \log \vert {\bf
x} - {\bf y} \vert = - {\partial \over \partial x^i} \phi ({\bf
x} - {\bf y}).$$
This means that Eq. (6.8) can be written
$${\bf A}({\bf x},t) = - {g \over 2\pi \kappa} \int d^2y
{\nabla}_{\bf x} \phi ({\bf x} - {\bf y}) \rho ({\bf y},t).
\eqno(6.10)$$
If I am allowed to move the derivation operator outside the
integral (this is not trivial: $\phi$ is not single--valued; see
{\cite{13}} for a discussion about the validity of this step),
(6.10) is a pure gauge ${\bf A} = \nabla \Lambda$, with
$$\Lambda ({\bf x},t) = - {g \over 2 \pi \kappa} \int d^2y \phi
({\bf x} - {\bf y}) \rho ({\bf y},t). \eqno(6.11)$$

Returning to Eq. (6.4), now for $\alpha = i = 1,2$:
$$F_{i0} = {\partial}_i A_0 - {\partial}_0 A_i = {g \over
\kappa} {\epsilon}_{ik}j^k, \eqno(6.12)$$
we can, using the continuity equation
$${\partial \rho \over \partial t} + \nabla \cdot {\bf j} = 0,
\eqno(6.13)$$
solve also for $A_0$, with the result
$$A_0({\bf x},t) = {g \over \kappa} \int d^2y G({\bf x} -{\bf
y}) {\epsilon}^{ik} {\partial j^k \over \partial y^i}.
\eqno(6.14)$$
Upon integrating by parts and using Eq. (6.13) again, we arrive
at
$$A_0({\bf x},t) = {\partial \over \partial t} {g \over 2 \pi
\kappa} \int d^2y \phi ({\bf x} - {\bf y}) \rho ({\bf y},t) = -
{\partial \over \partial t} \Lambda ({\bf x},t). \eqno(6.15)$$

Thus we have shown that $A_{\mu} = -{\partial}_{\mu}\Lambda$, a
pure gauge. A gauge transformation will remove it and bring
(6.1) to the form
$${\cal L'}_0 = i{\tilde \psi}^{\dagger} {\partial}_0 {\tilde
\psi} + {1 \over 2m} {\tilde \psi}^{\dagger} {\nabla}^2 {\tilde
\psi}, \eqno(6.16)$$
with the transformed fields
$${\tilde \psi}({\bf x},t) = e^{-ig\Lambda({\bf x},t)} \psi
({\bf x},t)$$
$${\tilde \psi}^{\dagger}({\bf x},t) = {\psi}^{\dagger}({\bf
x},t) e^{ig\Lambda({\bf x},t)}. \eqno(6.17)$$
The gauge parameter $\Lambda$ is an operator, and this will
affect the algebra of the ${\tilde \psi}$--operators. $\psi$ and
${\psi}^{\dagger}$ were ordinary boson fields satisfying the
algebra (2.7), (2.8). From Eq. (6.11) we deduce
$$ \lbrack \psi ({\bf x},t), \Lambda ({\bf y},t) \rbrack_{-} = -
{g \over 2\pi \kappa} \phi ({\bf y} - {\bf x}) \psi ({\bf x},t).
\eqno(6.18)$$
with this result, it is straightforward to derive the algebra of
the transformed operators (6.17). We get, using the notation
(5.14), e.g.
$${\tilde \psi}({\bf x},t){\tilde \psi}({\bf y},t) = e^{i\nu
(\phi({\bf y} - {\bf x}) - \phi({\bf x} - {\bf y}))} {\tilde
\psi}({\bf y},t) {\tilde \psi}({\bf x},t). \eqno(6.19)$$
This is of the general form (2.13), but we have to ask is this
really what we want?

All hinges on the argument funtion $\phi$ appearing in (6.19).
It is a multivalued function a priori, but can be made single--
valued by introducing cuts across which $\phi$ jumps by $2\pi$.
Taking the cut in the direction of the positive $x^1$-axis, so
that $\phi({\bf e}_1 + \epsilon {\bf e}_2) \rightarrow 0,
\phi({\bf e}_1 - \epsilon {\bf e}_2) \rightarrow 2\pi$ as
$\epsilon \rightarrow 0$, we have {\cite{18}}
\begin{eqnarray*}
\phi({\bf y} - {\bf x}) - \phi({\bf x} - {\bf y}) & = &
\pi {\rm sgn}(x^2-y^2), x^2 \not= y^2;\\
 & = & \pi {\rm sgn}(x^1- y^1), x^2=y^2.
\end{eqnarray*}
Cutting along the negative $x^1$-axis gives the opposite values
for the difference of the arguments.

In either case, given the locations ${\bf x}$ and ${\bf y}$, the
phase factor in (6.19) takes a specific value, either
$e^{i\nu\pi}$ or $e^{-i\nu\pi}$. Note that this is the phase
acquired by the wave function for two anyons in a state $\vert
\Phi \rangle$ :
$$\Psi ({\bf x},{\bf y}) = \langle 0 \vert {\tilde \psi}({\bf
x}) {\tilde \psi}({\bf y}) \vert \Phi \rangle \eqno(6.20)$$
under an interchange of the arguments. But this is not enough.
The phase of the wave function should be able to change {\it
both} by $+\pi\nu$ and $-\pi\nu$ in response to which way we
braid in interchanging ${\bf x}$ and ${\bf y}$. In other words,
states built by the creation operators ${\tilde \psi}^{\dagger}$
acting on the vacuum, with a specific, single--valued choice of
the argument function $\phi ({\bf x})$, do not provide a
representation of the braid group $B_N$.

It is evident, however, that this is the best we can achieve
with {\it local} operators ${\tilde \psi}, {\tilde
\psi}^{\dagger}$. Local information, in the meaning of initial
and final positions of particles, is simply not sufficient to
code the braiding, where we have to specify also which way
around each other the particles passed in interchanging
positions.

The only solution to this dilemma (and in fairness it should be
added that not all experts see this as any dilemma) within the
present framework, is to give the argument function $\phi$ a
definition that makes ${\tilde \psi}$ a {\it nonlocal} operator.
This point has been emphasized by Semenoff {\cite{12}}.

Let us return to the transformed field operators (6.17) with the
operator--valued gauge function (6.11). We give the following
definition of the argument function $\phi$ in (6.11): To the
point ${\bf x}$ we attach a curve $C_{\bf x}$ starting from some
reference point ${\bf P}$ (infinity is a convenient choice) and
ending at ${\bf x}$. Let ${\bf x'}$ move along $C_{\bf x}$ from
${\bf P}$ to ${\bf x}$. $\phi_{C_{\bf x}}({\bf x},{\bf y})$ is
then defined as the change in the polar angle of ${\bf x'}$, as
seen from ${\bf y}$, as ${\bf x'}$ moves from ${\bf P}$ to ${\bf
x}$ along $C_{\bf x}$:
$$\phi_{C_{\bf x}}({\bf x},{\bf y}) = \int_{C_{\bf x}} d{\bf l}
\cdot \nabla_l \phi ({\bf l} - {\bf y}). \eqno(6.21)$$

The corresponding transformed field operators now depend on the
curve $C_{\bf x}$ as well:
$${\tilde \psi}[C_{\bf x}] = e^{i\nu \int d^2y\:\phi_{C_{\bf
x}}({\bf x},{\bf y}) \rho ({\bf y})} \psi ({\bf x})$$
$${\tilde \psi}^{\dagger} [C_{\bf x}] = {\psi}^{\dagger}({\bf
x}) e^{- i\nu \int d^2y\:\phi_{C_{\bf x}}({\bf x},{\bf y}) \rho
({\bf y})}. \eqno(6.22)$$
A state of two localized anyons is now given by
$${\tilde \psi}^{\dagger}[C_{\bf x}] {\tilde \psi}^{\dagger}
[C_{\bf y}] \vert 0 \rangle = e^{-i\nu{\phi}_{C_{\bf x}} ({\bf
x}, {\bf y})} \vert {\bf x}, {\bf y} \rangle , \eqno(6.23)$$
where $ \vert {\bf x}, {\bf y} \rangle$ is the symmetric two-
boson state
$$\vert {\bf x}, {\bf y} \rangle = {\psi}^{\dagger}({\bf x})
{\psi}^{\dagger}({\bf y}) \vert 0 \rangle. \eqno(6.24)$$

The action of an operator $U[\sigma]$ implementing the braiding
transformation $\sigma$ is taken to mean extending the curves
$C_{\bf x} , C_{\bf y}$ by pieces describing the braiding (e.g.
adding half a circle, clockwise or counterclockwise oriented and
centered on ${\bf y}$, to $C_{\bf x}$ and then straight line
pieces to both $C_{\bf x}$ and $C_{\bf y}$ so that their
endpoints are interchanged). It is clear from Eq. (6.23) that
this gives the two-anyon state just the correct phase
${\chi}^*(\sigma)$ corresponding to the braiding $\sigma$. In
this way we have a means of representing all of $B_N$, albeit
with nonlocal states.

\section{Epilogue}
We have failed miserably in attaining our initial goal of
finding a simple, computationally useful formulation of the
many-anyon problem, like the second quantization of bosons and
fermions. The solution we ended up with is complicated and not
particularly workable in practical calculations. Some authors
even maintain that Chern--Simons field theory has nothing to do
with anyons, e.g. {\cite{19}}.

The rather formal manipulations we carried out in the previous
section can be put on a more rigorous footing by formulating the
theory on a lattice. This is especially true of the relativistic
theory, which presents difficulties of its own. Again, the
theory with a Maxwell term is better behaved {\cite{20,21}}, the
pure Chern--Simons theory presents peculiar pitfalls
{\cite{22}}. (It is interesting to note that Fr{\"o}hlich and
Marchetti {\cite{21}} formulate anyon statistics for asymptotic
states in p-space --- probably the right thing to do in
relativistic theories!)

But the conclusions remain the same: Anyons are described by
nonlocal operators, which are hard to deal with. If we insist on
a local formulation, we have to hide the statistics in an
interaction with a Chern--Simons field, and pay the price of
handling the extra interaction through whatever means are
avilable.

For lack of space, time and competence many topics of anyon
physics have not been treated in these lectures. Among these
are: The relation between spin and statistics, the fractional
quantum Hall effect and the statistical mechanics of anyons. For
these, I have to refer you to the reviews already cited
{\cite{5,11,12}}, to which a few more can be added {\cite{23}}.

\end{document}